\documentclass[a4paper,12pt]{article}
\usepackage{amsfonts}

\usepackage{amsmath}

\newtheorem{theorem}{Theorem}

\newtheorem{proposition}[theorem]{Proposition}
\newtheorem{remark}[theorem]{Remark}

\newenvironment{proof}[1][Proof]{\textbf{#1.} }{\ \rule{0.5em}{0.5em}}

\begin{document}

\author{G. Cassinelli\thanks{G.~Cassinelli,
Dipartimento di Fisica, Universit\`a di Genova, and I.N.F.N.,
Sezione di Genova, Via Dodecaneso~33, 16146 Genova, Italy. e-mail:
cassinelli@ge.infn.it}, E. De Vito\thanks{E.~De Vito,
Dipartimento di Matematica, Universit\`a di Modena e Reggio
Emilia, Via Campi 213/B, 41100 Modena, Italy, and I.N.F.N., Sezione di
Genova, Via Dodecaneso~33, 16146 Genova, Italy. e-mail: devito@unimo.it}, A.
Toigo\thanks{A.~Toigo,
Dipartimento di Fisica, Universit\`a di Genova, and I.N.F.N.,
Sezione di Genova, Via Dodecaneso~33, 16146 Genova, Italy. e-mail:
toigo@ge.infn.it}}
\title{Positive operator valued measures covariant with respect to an irreducible representation}
\date{February 26, 2003}
\maketitle

\begin{abstract}
Given an irreducible representation of a group $G$, we show that all the
covariant positive operator valued measures based on $G/Z$, where $Z$ is a
central subgroup, are described by trace class, trace one positive operators.
\end{abstract}

\setlength\arraycolsep{2pt}

\section{Introduction}

It is well known~\cite{Isot,scud} that, given a square-integrable
representation $\pi$ of a unimodular group $G$ and a trace class, trace
one positive operator $T$, the family of operators 
\begin{equation*}
Q(X)=\int_X \pi(g)T\pi(g^{-1})d\mu_G(g),
\end{equation*}
defines a positive operator valued
measure (POVM) on $G$ covariant with respect to $\pi$ 
($\mu_G$ is a Haar measure on $G$). In this paper, we prove that
all the covariant POVMs are of the above form for some $T$. More precisely,
we show this result for non-unimodular groups and for POVMs based on the
quotient space $G/Z$, where $Z$ is a central subgroup.

Let $G$ be a locally compact second countable topological group and $Z$ be a central
closed subgroup. We denote by $G/Z$ the quotient group and by $\dot{g}\in
G/Z $ the equivalence class of $g\in G$. If $a\in G$ and $\dot{g}\in G/Z$,
we let $a\left[ \dot{g}\right] =\dot{a}\dot{g}$ be the natural action of $a$
on the point $\dot{g}$.

Let $\mathcal{B}\left( G/Z\right) $ be the Borel $\sigma $-algebra of $G/Z$.
We fix a left Haar measure $\mu _{G/Z}$ on $G/Z$. Moreover, we denote by $\Delta $
the modular function of $G$ and of $G/Z$.

By \emph{representation} we mean a strongly continuous unitary
representation of $G$ acting on a complex and separable Hilbert space,
with scalar product $\left\langle \cdot ,\cdot \right\rangle $ linear in the
first argument.

Let $\left( \pi ,\mathcal{H}\right) $ be a representation of $G$. A positive
operator valued measure $Q$ defined on $G/Z$ and such that

\begin{enumerate}
\item  $Q\left( G/Z\right) =I$;

\item  for all $X\in \mathcal{B}\left( G/Z\right) $, 
\begin{equation*}
\pi \left( g\right) Q\left( X\right) \pi \left( g^{-1}\right) =Q\left( g
\left[ X\right] \right) \qquad \forall g\in G
\end{equation*}
\end{enumerate}

\noindent is called $\pi $-\emph{covariant} POVM on $G/Z$.

Given a representation $\left( \sigma ,\mathcal{K}\right) $ of $Z$, we
denote by $\left( \lambda^{\sigma} , P^{\sigma} , \mathcal{H}^{\sigma}
\right)$ the imprimitivity system unitarily induced by $\sigma$. We recall
that $\mathcal{H}^{\sigma}$ is the Hilbert space of ($\mu _{G}$-equivalence
classes of) functions $f:G\longrightarrow \mathcal{K}$ such that

\begin{enumerate}
\item  $f$ is weakly measurable;

\item  for all $z\in Z$, 
\begin{equation*}
f\left( gz\right) =\sigma \left( z^{-1}\right) f\left( g\right) \qquad
\forall g\in G\text{;}
\end{equation*}

\item  
\begin{equation*}
\int_{G/Z}\left\| f\left( g\right) \right\| _{\mathcal{K}}^{2}\,\text{d}\mu
_{G/Z}\left( \dot{g}\right) <+\infty
\end{equation*}
\end{enumerate}

\noindent with scalar product 
\begin{equation*}
\left\langle f_{1},f_{2}\right\rangle _{\mathcal{H}^{\sigma
}}=\int_{G/Z}\left\langle f_{1}\left( g\right) ,f_{2}\left( g\right)
\right\rangle _{\mathcal{K}}\text{d}\mu _{G/Z}\left( \dot{g}\right) \text{.}
\end{equation*}
The representation $\lambda^\sigma$ acts on $\mathcal{H}^{\sigma }$ as 
\begin{equation*}
\left( \lambda ^{\sigma }\left( a\right) f\right) \left( g\right) :=f\left(
a^{-1}g\right) \qquad g\in G
\end{equation*}
for all $a\in G$. The projection valued measure $P^{\sigma }$ is given by 
\begin{equation*}
\left( P^{\sigma }\left( X\right) f\right) \left( g\right) :=\chi _{X}\left( 
\dot{g}\right) f\left( g\right) \qquad g\in G\text{.}
\end{equation*}
for all $X\in \mathcal{B}\left( G/Z\right) $, where $\chi_X $ is the characteristic
function of the set $X$.

We recall some basic properties of square integrable representations modulo
a central subgroup. We refer to Ref.~\cite{Borel} for $G$ unimodular
and $Z$ arbitrary and to Ref.~\cite{df} for $G$ non-unimodular and
$Z=\{e\}$. Combining these proofs, one obtains the
following result.

\begin{proposition}
\label{rivaldo} Let $\left( \pi ,\mathcal{H}\right) $ be an irreducible
representation of $G$ and $\gamma $ be the character of $Z$ such that 
\begin{equation*}
\pi \left( z\right) =\gamma \left( z\right) I_{\mathcal{H}}\qquad \forall
z\in Z\text{.}
\end{equation*}
The following facts are equivalent:

\begin{enumerate}
\item  there exists a vector $u\in \mathcal{H}$ such that 
\begin{equation}
0<\int_{G/Z}\left| \left\langle u,\pi \left( g\right) u\right\rangle _{
\mathcal{H}}\right| ^{2}\text{d}\mu _{G/Z}\left( \dot{g}\right) <+\infty 
\text{;}  \label{galliani}
\end{equation}

\item  $\left( \pi ,\mathcal{H}\right) $ is a subrepresentation of $\left(
\lambda ^{\gamma },\mathcal{H}^{\gamma }\right) $.
\end{enumerate}

If any of the above conditions is satisfied, there exists a selfadjoint
injective positive operator $C$ with dense range such that 
\begin{equation*}
\pi \left( g\right) C=\Delta \left( g\right) ^{-\frac{1}{2}}C\pi \left(
g\right) \qquad \forall g\in G\text{,}
\end{equation*}
and an isometry $\Sigma :\mathcal{H}\otimes \mathcal{H}^{\ast }\rightarrow 
\mathcal{H}^{\gamma }$ such that

\begin{enumerate}
\item  for all $u\in \mathcal{H}$ and $v\in \mathop{\rm{dom}}\nolimits C$ 
\begin{equation*}
\Sigma (u\otimes v^{\ast })(g)=\left\langle u,\pi \left( g\right)
Cv\right\rangle _{\mathcal{H}}\qquad g\in G\text{,}
\end{equation*}

\item  for all $g\in G$ 
\begin{equation*}
\Sigma (\pi (g)\otimes I_{\mathcal{H}^{\ast }})=\lambda (g)\Sigma \text{,}
\end{equation*}

\item  the range of $\Sigma $ is the isotypic space of $\pi $ in
$\mathcal{H}^{\gamma }$.
\end{enumerate}
\end{proposition}

If Eq.~(\ref{galliani}) is satisfied, $\left( \pi ,\mathcal{H}\right) $ is
called \emph{square-integrable modulo} $Z$. The square root of $C$ is
called \emph{formal degree} of $\pi$ (see Ref.~\cite{df}).
In particular, when $G$ is unimodular, $C$ is
 a multiple of the identity.

\section{Characterization of $Q$}

We fix an irreducible representation $\left( \pi ,\mathcal{H}\right) $ of $G$
and let $\gamma$ be the character such that $\left. \pi \right| _{Z}=\gamma
I_{\mathcal{H}}$. The following theorem characterizes all the POVM on $G/Z$
covariant with respect to $\pi $ in terms of positive trace one operators on 
$\mathcal{H}$.

\begin{theorem}
\label{Teo. centr.}The irreducible representation $\pi $ admits a covariant
POVM based on $G/Z$ if and only if $\pi$ is square-integrable modulo $Z$. 

In
this case, let $C$ be the square root of the formal degree of $\pi$.
There exists a one-to-one correspondence between covariant POVMs $
Q$ on $G/Z$ and positive trace one operators $T$ on $\mathcal{H}$ 
given by 
\begin{equation}
\left\langle Q_{T}\left( X\right) v,u\right\rangle _{\mathcal{H}}
=\int_{X}\left\langle TC\pi \left( g^{-1}\right) v,C\pi \left(
g^{-1}\right) u\right\rangle _{\mathcal{H}}\text{d}\mu _{G/Z}\left( \dot{g}
\right)   \label{La Povm}
\end{equation}
for all $u,v\in \mathop{\rm{dom}}\nolimits C$ and $X\in \mathcal{B}\left( G/Z\right) $.
\end{theorem}

\begin{proof}
Let $Q$ be a $\pi $-covariant POVM. According to the generalized
imprimitivity theorem~\cite{Catt} there exists a representation $\left(
\sigma ,\mathcal{K}\right) $ of $Z$ and an isometry $W:\mathcal{H}
\longrightarrow \mathcal{H}^{\sigma }$ intertwining $\pi $ with $\lambda
^{\sigma }$ such that 
\begin{equation*}
Q\left( X\right) =W^{\ast }P^{\sigma }\left( X\right) W
\end{equation*}
for all $X\in \mathcal{B}\left( G/Z\right) $.

Define the following closed invariant subspace of $\mathcal{K}$ 
\begin{equation*}
\mathcal{K}_{\gamma }=\left\{ v\in \mathcal{K}\mid \sigma \left( z\right)
v=\gamma \left( z\right) v\right\} \text{.}
\end{equation*}
Let $\sigma _{1}$ and $\sigma _{2}$ be the restrictions of $\sigma $ to
$\mathcal{K}_{\gamma }$ and $\mathcal{K}_{\gamma }^{\perp }$ respectively.
The induced imprimitivity system $\left( \lambda ^{\sigma },P^{\sigma
},\mathcal{H}^\sigma\right) $ decomposes into the orthogonal sum 
\begin{equation*}
\mathcal{H}^{\sigma }=\mathcal{H}^{\sigma _{1}}\oplus \mathcal{H}^{\sigma
_{2}}\text{.}
\end{equation*}
If $f\in \mathcal{H}^{\sigma }$ and $z\in Z$, then 
\begin{equation*}
\left( \lambda ^{\sigma }\left( z\right) f\right) \left( g\right) =f\left(
z^{-1}g\right) =f\left( gz^{-1}\right) =\sigma \left( z\right) f\left(
g\right) \qquad g\in G\text{.}
\end{equation*}
On the other hand, if $u\in \mathcal{H}$ and $z\in Z$, we have 
\begin{equation*}
\left( \lambda ^{\sigma }\left( z\right) Wu\right) \left( g\right) =\left(
W\pi \left( z\right) u\right) \left( g\right) =\gamma \left( z\right) \left(
Wu\right) \left( g\right) \qquad g\in G\text{.}
\end{equation*}
It follows that $\left( Wu\right) \left( g\right) \in \mathcal{K}_{\gamma }$
for $\mu _{G}$-almost every $g\in G$, that is, $Wu\in \mathcal{H}^{\sigma
_{1}}$. So it is not restrictive to assume that 
\begin{equation*}
\sigma =\gamma I_{\mathcal{K}}
\end{equation*}
for some Hilbert space $\mathcal{K}$. Clearly, we have 
\begin{equation*}
\mathcal{H}^{\sigma }=\mathcal{H}^{\gamma }\otimes \mathcal{K}\text{,\qquad }
\lambda ^{\sigma }=\lambda ^{\gamma }\otimes I_{\mathcal{K}}\text{.}
\end{equation*}
In particular, $\pi $ is a subrepresentation of $\lambda ^{\gamma }$, hence
it is square-integrable modulo $Z$.

Due to Prop.~\ref{rivaldo}, the operator $W^{\prime }=\left( \Sigma ^{\ast
}\otimes I_{\mathcal{K}} \right) W$ is an isometry from
$\mathcal{H}$ to $\mathcal{H} \otimes \mathcal{H}^{\ast }\otimes \mathcal{K}$
such that 
\begin{equation*}
W^{\prime }\pi (g)=\pi (g)\otimes I_{\mathcal{H}^{\ast }\otimes \mathcal{K}
}\qquad g\in G.
\end{equation*}
Since $\pi $ is irreducible, there is a unit vector $B\in  
\mathcal{H}^{\ast} \otimes \mathcal{K}$ such that 
\begin{equation*}
W^{\prime }u=u\otimes B\qquad \forall u\in \mathcal{H}.
\end{equation*}
Let $(e_{i})_{i\geq 1}$ be an orthonormal basis of $\mathcal{H}$ such that
$e_{i}\in \mathop{\rm{dom}}\nolimits C$, then 
\begin{equation*}
B=\sum e_{i}^{\ast }\otimes k_{i},
\end{equation*}
where $k_{i}\in \mathcal{K}$ and $\sum_{i}\left\| k_{i}\right\| _{\mathcal{K}}^{2}
=1$.

If $u\in \mathop{\rm{dom}}\nolimits C$, one has that 
\begin{eqnarray*}
(Wu)(g) &=&\left[ (\Sigma \otimes I_{\mathcal{K}})\left( u\otimes B\right)
\right] (g) \\
&=&\sum\nolimits_{i}\Sigma (u\otimes e_{i}^{\ast })(g)\otimes k_{i} \\
&=&\sum\nolimits_{i}\left\langle u,\pi \left( g\right) Ce_{i}
\right\rangle_{\mathcal{H}}\otimes k_{i} \\
&=&\sum\nolimits_{i}\left\langle C\pi \left( g^{-1}\right)
u,e_{i}\right\rangle _{\mathcal{H}}\otimes k_{i} \\
&=&\sum\nolimits_{i}(e_{i}^{\ast }\otimes k_{i})(C\pi \left( g^{-1}\right)
u),
\end{eqnarray*}
where the series converges in $\mathcal{H}^{\sigma }$. On the other hand,
for all $g\in G$ the series $\sum_{i}(e_{i}^{\ast }\otimes k_{i})(C\pi
\left( g^{-1}\right) u)$ converges to $BC\pi \left( g^{-1}\right) u$, where we identify 
$\mathcal{H}^\ast\otimes\mathcal{K}$ with the space of Hilbert-Schmidt operators. 
By unicity of the limit 
\begin{equation*}
(Wu)(g)=BC\pi \left( g^{-1}\right) u\qquad g\in G.
\end{equation*}

If $u,v\in \mathop{\rm{dom}}\nolimits C$, the corresponding covariant POVM is
given by 
\begin{eqnarray*}
\left\langle Q\left( X\right) v,u\right\rangle _{\mathcal{H}}
&=& \left\langle P^{\sigma }\left( X\right) Wv,Wu
\right\rangle_{\mathcal{H}^{\sigma }} \\
&=&\int_{G/Z}\chi _{X}\left( \dot{g}\right) \left\langle BC\pi \left(
g^{-1}\right) v,BC\pi \left( g^{-1}\right) u\right\rangle _{\mathcal{H}}
\text{d}\mu _{G/Z}\left( \dot{g}\right)  \\
&=&\int_{X}\left\langle T C\pi \left( g^{-1}\right) v,C\pi \left(
g^{-1}\right) u\right\rangle _{\mathcal{H}}\text{d}\mu _{G/Z}\left( \dot{g}
\right) \text{,}
\end{eqnarray*}
where 
\begin{equation*}
T :=B^{\ast }B
\end{equation*}
is a positive trace class trace one operator on $\mathcal{H}$.

Conversely, assume that $\pi$ is square integrable and let $T$ be
a positive trace class trace one operator on $\mathcal{H}$. Then 
\begin{equation*}
B:=\sqrt{T}
\end{equation*}
is a (positive) operator belonging to $\mathcal{H}^{\ast }\otimes \mathcal{H}
$ such that $B^{\ast }B=T$ and $\left\| B\right\| _{\mathcal{H}^{\ast
}\otimes \mathcal{H}}=1$. The operator $W$ defined by 
\begin{equation*}
Wv:=\left( \Sigma \otimes I_{\mathcal{H}}\right) \left( v\otimes B\right)
\qquad \forall v\in \mathcal{H}
\end{equation*}
is an isometry intertwining $\left( \pi ,\mathcal{H}\right) $ with the
representation $\left( \lambda ^{\sigma },\mathcal{H}^{\sigma }\right) $,
where 
\begin{equation*}
\sigma =\gamma I_{\mathcal{H}}\text{.}
\end{equation*}
Define $Q_T$ by
\begin{equation*}
Q_{T}\left( X\right) =  W^{\ast }P^{\sigma }\left( X\right) W \qquad X\in\mathcal{B}(G/Z)\text{.}
\end{equation*}
With the same computation as above, one has that 
\begin{equation*}
\left\langle Q_{T}\left( X\right)u,v\right\rangle _{\mathcal{H}} = 
\int_{X}\left\langle TC\pi \left( g^{-1}\right) v,C\pi \left(g^{-1}\right) u
\right\rangle _{\mathcal{H}}\text{d}\mu _{G/Z}\left( \dot{g}
\right) 
\end{equation*}
for all $u,v\in \mathop{\rm{dom}}\nolimits C$.

Finally, we show that the correspondence $T\longmapsto Q_{T}$ is injective.
Let $T_{1}$ and $T_{2}$ be positive trace one operators on $\mathcal{H}$,
with $Q_{T_{1}}=Q_{T_{2}}$. Set $T=T_{1}-T_{2}$. Since $\pi $ is strongly
continuous, for all $u,v\in \mathop{\rm{dom}}\nolimits C$ the map
{\setlength\arraycolsep{0pt}
\begin{eqnarray*}
&& G/Z\ni \dot{g}\longmapsto \left\langle TC\pi \left( g^{-1}\right) v,C\pi
\left( g^{-1}\right) u\right\rangle _{\mathcal{H}}  \\
&&\quad \quad \quad \quad \quad = \Delta (\dot{g})^{-1}\left\langle
T\pi \left( g^{-1}\right) Cv,\pi 
\left( g^{-1}\right) Cu\right\rangle _{\mathcal{H}}
\in \mathbb{C}
\end{eqnarray*}
}is continuous. Since 
\begin{equation*}
\int_{X}\left\langle TC\pi \left( g^{-1}\right) v,C\pi \left( g^{-1}\right)
u\right\rangle _{\mathcal{H}}\text{d}\mu _{G/Z}\left( \dot{g}\right)
=\left\langle \left[ Q_{T_{1}}\left( X\right) -Q_{T_{2}}\left( X\right)
\right] v,u\right\rangle _{\mathcal{H}}=0
\end{equation*}
for all $X\in \mathcal{B}\left( G/Z\right) $, we have 
\begin{equation*}
\left\langle TC\pi \left( g^{-1}\right) v,C\pi \left( g^{-1}\right)
u\right\rangle _{\mathcal{H}}=0\qquad \forall \dot{g}\in G/Z\text{.}
\end{equation*}
In particular, 
\begin{equation*}
\left\langle TCv,Cu\right\rangle _{\mathcal{H}}=0\text{,}
\end{equation*}
so that, since $C$ has dense range, $T=0$.
\end{proof}

\begin{remark}
Scutaru shows in Ref.~$\cite{scud}$ that there exists a one-to-one
correspondence between positive trace one operators on $\mathcal{H}$ and
covariant POVMs $Q$ based on $G/Z$ with the property 
\begin{equation}
\mathop{\rm{tr}}\nolimits Q\left( K\right) <+\infty \label{inzaghi}
\end{equation}
for all compact sets $K\subset G/Z$. Theorem $\ref{Teo. centr.}$ shows that 
\emph{every} covariant POVM $Q$ based on $G/Z$ shares property $(\ref{inzaghi})$.
\end{remark}


\begin{remark}
If $G$ is unimodular, then $K=\lambda I$, with $\lambda >0$, and one can normalize 
$\mu _{G/Z}$  so that $\lambda =1$. Hence,
\begin{equation*}
Q_{T}\left( X\right) =\int_{X}\pi \left( g\right) T\pi \left( g^{-1}\right) 
\text{d}\mu _{G/Z}\left( \dot{g}\right) \qquad \forall X\in \mathcal{B}
\left( G/Z\right) \text{,}
\end{equation*}
the integral being understood in the weak sense.
\end{remark}

\begin{remark}
If $T=\eta^{\ast }\otimes \eta$, with $\eta \in \mathop{\rm{dom}}\nolimits C$
and $\left\| \eta\right\|_{\mathcal{H}}
=1$, we observe that 
\begin{eqnarray*}
\left\langle Q_{T}\left( X\right) v,u\right\rangle _{\mathcal{H}}
&=&\int_{X}\left\langle C\pi \left( g^{-1}\right) v,\eta\right\rangle _{
\mathcal{H}}\left\langle \eta,C\pi \left( g^{-1}\right) u\right\rangle _{
\mathcal{H}}\text{d}\mu _{G/Z}\left( \dot{g}\right)  \\
&=&\int_{X}\left\langle v,\pi \left( g\right) C\eta\right\rangle _{\mathcal{H}
}\left\langle \pi \left( g\right) C\eta,u\right\rangle _{\mathcal{H}}\text{d}
\mu _{G/Z}\left( \dot{g}\right)  \\
&=&\int_{X}\left( W_{C\eta}v\right) \left( g\right) \overline{\left(
W_{C\eta}u\right) \left( g\right) }\text{d}\mu _{G/Z}\left( \dot{g}\right) 
\end{eqnarray*}
for all $u,v\in \mathop{\rm{dom}}\nolimits C$, where $W_{C\eta}:\mathcal{H}\longrightarrow 
\mathcal{H}^{\gamma }$ is the wavelet operator associated to the vector $C\eta$.
In particular, 
\begin{equation*}
Q_{T}\left( X\right) =W_{C\eta}^{\ast} P^{\gamma }\left( X\right)
W_{C\eta} \text{.}
\end{equation*}
\end{remark}

\section{Two examples}

\subsection{The Heisenberg group}

The Heisenberg group $H$ is $\mathbb{R}^{3}$ with
composition law
\begin{equation*}
\left( p,q,t\right) \left( p^{\prime },q^{\prime },t^{\prime }\right)
=\left( p+p^{\prime },q+q^{\prime },t+t^{\prime }+\frac{pq^{\prime
}-qp^{\prime }}{2}\right) \text{.}
\end{equation*}
The centre of $H$ is 
\begin{equation*}
Z=\left\{ \left( 0,0,t\right) \mid t\in \mathbb{R}\right\} \text{,}
\end{equation*}
and the quotient group $G/Z$ is isomorphic to the Abelian group $\mathbb{R}
^{2}$, with projection 
\begin{equation*}
q\left( p,q,t\right) =\left( p,q\right) \text{.}
\end{equation*}
The Heisenberg group is unimodular with Haar measure
\begin{equation*}
\text{d}\mu _{G/Z}\left( p,q\right) =\frac{1}{2\pi }\text{d}p\text{d}q \text{.}
\end{equation*}

Given an infinite dimensional Hilbert space $\mathcal{H}$ and an
orthonormal basis $(e_{n})_{n\geq 1}$, let $a$, $a^{\ast }$ be
the corresponding ladder operators. Define
\begin{eqnarray*}
Q &=&\frac{1}{\sqrt{2}}(a+a^{\ast }) \\
P &=&\frac{1}{\sqrt{2}i}(a-a^{\ast })
\end{eqnarray*}
It is known~\cite{Isot,Foll} that the representation
\begin{equation*}
\pi (p,q,t)=e^{i(t+pQ+qP)}
\end{equation*}
is square-integrable modulo $Z$ and $C=1$.

It follows from Theorem \ref{Teo. centr.} that \emph{any} $\pi$
-covariant POVM $Q$ based on $\mathbb{R}^2$ is of the form
\begin{equation*}
Q \left( X\right) = \frac{1}{2\pi }\int_{X} e^{i(pQ +qP)}T e^{-i(pQ +qP)}
\text{d}p \text{d}q \qquad X\in \mathcal{B}\left( \mathbb{R}^2\right) 
\end{equation*}
for some positive trace one operator on $\mathcal{H}$.
Up to our knowledge, the complete classification of the POVMs on
$\mathbb{R}^2$ covariant
 with respect to the Heisenberg group has been an
open problem till now.
 \subsection{The $ax+b$ group}

The $ax+b$ group is the semidirect product $G=\mathbb{R}\times ^{\prime }\mathbb{R}_{+}$,
where we regard $\mathbb{R}$  as additive group and $\mathbb{R}_{+}$ as multiplicative group. 
The composition law is 
\begin{equation*}
\left( b,a\right) \left( b^{\prime },a^{\prime }\right) =\left( b+ab^{\prime
},aa^{\prime }\right) \text{.}
\end{equation*}
The group $G$ is nonunimodular with left Haar measure
\begin{equation*}
\text{d}\mu _{G}\left( b,a\right) =a^{-2}\text{d}b\text{d}a
\end{equation*}
and modular function
\begin{equation*}
\Delta \left( b,a\right) =\frac{1}{a}\text{.}
\end{equation*}

Let $\mathcal{H}=L^{2}\left( \left( 0,+\infty \right) ,
\text{d}x\right) $ and $(\pi^{+}, \mathcal{H})$ be the representation of $G$ given by
\begin{equation*}
\left[ \pi ^{+}\left( b,a\right) f\right] \left( x\right) =a^{\frac{1}{2}}
e^{2\pi ibx}f\left( ax\right) \qquad x\in \left( 0,+\infty \right) \text{.}
\end{equation*}
It is known~\cite{Foll} that $\pi$ is square-integrable, and the square
root of
 its formal degree is
\begin{equation*}
\left( Cf\right) \left( x\right) =\Delta \left( 0,x\right) ^{\frac{1}{2}
}f\left( x\right) =x^{-\frac{1}{2}}f\left( x\right) \qquad x\in \left(
0,+\infty \right) 
\end{equation*}
acting on its natural domain.

By means of Theorem \ref{Teo. centr.} every POVM based on $
G$ and covariant with respect to $\pi^{+} $ is described by a positive
trace one operator $T$
according to Eq.~\ref{La Povm}. Explicitely, let $\left(
e_{i}\right) _{i\geq 1}$ be an orthonormal basis of eigenvectors of $T$ and
$\lambda _{i}\geq 0$ be the corresponding eigenvalues. If $u\in L^{2}\left(
\left( 0,+\infty \right) ,\text{d}x\right) $ is such that $x^{-\frac{1}{2}}
u\in L^{2}\left( \left( 0,+\infty \right) ,\text{d}x\right) $, the $\pi ^{+}
$-covariant POVM corresponding to $T$ is given by 
\begin{eqnarray*}
\left\langle Q_{T}\left( X\right) u,u\right\rangle _{\mathcal{H}}
&=&\int_{X}\left\langle TC\pi ^{+}\left( g^{-1}\right) u,C\pi ^{+}\left(
g^{-1}\right) u\right\rangle _{\mathcal{H}}\text{d}\mu _{G}\left( g\right) 
\\
&=&\int_{X}\sum\nolimits_{i}\lambda _{i}\left| \left\langle C\pi ^{+}\left(
g^{-1}\right) u,e_{i}\right\rangle _{\mathcal{H}}\right| ^{2}\text{d}\mu
_{G}\left( g\right)  \\
&=&\sum\nolimits_{i}\lambda _{i}\int_{X}\left| \int_{\mathbb{R}_{+}}
x^{-\frac{1}{2}}a^{-\frac{1}{2}}e^{-\frac{2\pi ibx}{a}}u\left(
\frac{x}{a}\right) 
 \overline{e_{i}\left( x\right) }\text{d}x\right|
^{2}a^{-2}\text{d}b\text{d}a
\text{.}
\end{eqnarray*}

\end{document}